\documentclass[preprint,aps,tightenlines,floatfix,showpacs,byrevtex]{revtex4}
\usepackage{epsfig}
\usepackage{graphics}
\usepackage{amsmath}
\newcommand{\be}{\begin{equation}}
\newcommand{\ee}{\end{equation}}

\begin{document}

\preprint{
\vbox{
\hbox{ADP-03-115/T553}
\hbox{JLAB-THY-03-30}
}}

\title{Spin-3/2 Nucleon and $\Delta$ Baryons in Lattice QCD}

\author{J.M.~Zanotti, 
        D.B.~Leinweber,
        A.G.~Williams and
        J.B.~Zhang                  \\
        (CSSM Lattice Collaboration)}

\affiliation{Department of Physics and Mathematical Physics, and\\
        Special Research Centre for the
        Subatomic Structure of Matter,                          \\
        University of Adelaide, 5005, Australia}
\author{W.~Melnitchouk}
\affiliation{Jefferson Lab, 12000 Jefferson Avenue,
        Newport News, VA 23606, USA}

\author{S.~Choe}
\affiliation{Korea Advanced Institute of Science and Technology,
373-1, Guseong-dong, Yuseong-gu, Daejon, Republic of Korea 305-701}

\begin{abstract}
We present first results for masses of spin-${3\over 2}$ $N$ and
$\Delta$ baryons in lattice QCD using Fat-Link Irrelevant Clover
(FLIC) fermions. 
Spin-${3\over 2}$ interpolating fields
providing overlap with both spin-$\frac{3}{2}$ and spin-${1\over 2}$
states are considered. 
In the isospin-${1\over 2}$ sector, we observe, after appropriate spin
and parity projection, a strong signal for the $J^P={3\over 2}^-$
state
together with a weak but discernible signal for the $\frac{3}{2}^+$ state
with a mass splitting near that observed experimentally.
We also find good agreement between the ${1\over 2}^\pm$ masses and earlier
nucleon mass simulations with the standard spin-${1\over 2}$ interpolating field.
For the isospin-${3\over 2}$ $\Delta$ states, clear mass splittings are
observed between the various ${1\over 2}^\pm$ and ${3\over 2}^\pm$
channels, with the calculated level orderings in good agreement with
those observed empirically. 
\end{abstract}

\vspace{3mm}

\maketitle

\section{Introduction}

The level orderings in the baryon spectrum and mass splittings between
excited baryon multiplets provide important clues to the underlying
dynamics governing inter-quark forces and the relevant effective
degrees of freedom at low energy \cite{ISGUR}.  Considerable insights
into these and other problems of spectroscopy have been gained from
QCD-inspired phenomenological models, however, many fundamental
questions about the origins of the empirical spectrum remain
controversial \cite{CR}.

The resolution of some of these issues may only be possible with the help
of calculations of the spectrum in lattice QCD --- currently the
only first-principles method able to determine hadron properties
directly from the fundamental quark and gluon theory.
Recent advances in computational capabilities and more efficient
algorithms have enabled the first dedicated lattice QCD simulations of
the excited states of the nucleon to be undertaken
\cite{DEREK,LL,LEE,DGR,UKQCD,BNL,NSTAR,Nakajima:2001js}.
Lattice studies of excited hadrons are possible because at the current
unphysically large quark masses and finite volumes used in the
simulations, most excited states are stable.
Contact with experiment can be made via extrapolations incorporating
the nonanalytic behaviour of chiral effective field theory \cite{RHO}.
These studies are timely as they complement the first results from
the high precision measurements of the $N^*$ spectrum at Jefferson Lab
\cite{CLAS}.

In a recent paper \cite{NSTAR} we presented first results for the excited
nucleon and spin-${1\over 2}$ hyperon spectra using the Fat-Link
Irrelevant Clover (FLIC) quark action \cite{FATJAMES} with
an ${\cal O}(a^2)$-improved gluon action.
The FLIC action minimizes the effect of renormalization
of action improvement terms and displays excellent scaling
properties \cite{FATJAMES}.
Clear mass splittings are observed for $J^P={1\over 2}^+$ and
${1\over 2}^-$ states, as well as evidence for the sensitivity to
hyperfine splittings of odd-parity states for different interpolating
fields used.
On the other hand, no evidence is seen for overlap of three-quark
interpolating fields with the Roper-like resonances or the lowest-lying
odd-parity SU(3) singlet state, the $\Lambda(1405)$.
In this paper we extend the analysis of Ref.~\cite{NSTAR} to the 
spin-${3\over 2}$ sector, and present first results using the FLIC action, in
both the isospin-${1\over 2}$ and ${3\over 2}$ channels.

Mass splittings between states within SU(3) quark-model multiplets
provide another important motivation for studying higher spin baryons.
Understanding the mass splitting between the
$N{1\over 2}^-(1535)$ and $N{3\over 2}^-(1520)$,
for instance, or between the
$\Delta{1\over 2}^-(1620)$ and $\Delta{3\over 2}^-(1700)$,
can help identify the important mechanisms associated with the hyperfine
interactions, or shed light on the spin-orbit force, which has been a
central mystery in spectroscopy \cite{SUMRULES}.
In valence quark models, the degeneracy between the $N{1\over 2}^-$ and
$N{3\over 2}^-$ can be broken by a tensor force associated with mixing
between the $N^2$ and $N^4$ representations of SU(3) \cite{CR}, although
this generally leaves the $N{3\over 2}^-$ at a higher energy than the
$N{1\over 2}^-$.
On the other hand, a spin-orbit force is necessary to split the
$\Delta{3\over 2}^-$ and $\Delta{1\over 2}^-$ states.
In the Goldstone boson exchange model \cite{GR}, both of these pairs of
states are degenerate.
Model-independent analyses in the large $N_c$ limit have found that these
mass splittings receive important contributions from operators that do not
have a simple quark model interpretation \cite{GOITY}, such as those
simultaneously coupling spin, isospin and orbital angular momentum.
Of course, the coefficients of the various operators in such an analysis
must be determined phenomenologically and guidance from lattice QCD is
essential.

Our lattice simulations are performed on the Orion computer cluster
dedicated to lattice QCD at the CSSM, University of Adelaide.
In the isospin-${3\over 2}$ sector, after applying suitable parity and
spin projections, we present the first results for the
$\Delta{1\over 2}^+$ and $\Delta{1\over 2}^-$ states, as well as the
$\Delta{3\over 2}^-$.
Our results for the $\Delta{3\over 2}^+$ are also in good agreement
with earlier simulations \cite{LDW}.
A significant advance of this work is the observation of a discernible
signal for the $\Delta{1\over 2}^\pm$ state, which yielded a weak
signal in earlier simulations \cite{DGR}.
The lowest excitation of the ground state, namely the
$\Delta{1\over 2}^-$, is found to have a mass $\sim 350$--400~MeV above
the $\Delta{3\over 2}^+$, with the $\Delta{3\over 2}^-$ slightly heavier.
The $\Delta{1\over 2}^+$ state is found to lie $\sim 100$--200~MeV
above these, although the signal becomes weak at smaller quark masses.
This level ordering is consistent with that observed in the empirical
mass spectrum.

In the spin-${3\over 2}$ nucleon sector, there is good agreement for the
spin-projected ${1\over 2}^+$ and ${1\over 2}^-$ states with earlier
nucleon mass calculations \cite{NSTAR} using the standard 
spin-${1\over 2}$ nucleon interpolating field.
Furthermore, we find a good signal for the $N{3\over 2}^\pm$ states, with
a mass difference of $\sim 300$~MeV between the spin-$\frac{3}{2}$
parity partners.
The $N{1\over 2}^-$ and $N{3\over 2}^-$ states are approximately degenerate
as observed experimentally.

In Section~\ref{spin32sec} we outline the basic elements of
formulating spin-${3\over 2}$ baryons on the lattice, including the choice of
interpolating fields and projection operators.
A brief preliminary report of states using the formalism developed and
presented here 
appeared in Ref.~\cite{LeeSpin32}.
These results also supersede preliminary results reported in
Refs.~\cite{QNP,Latt02}.
In Section~\ref{spin32results}, our results using the FLIC action on a
large lattice volume at a fine lattice spacing represent the first
quantitative analysis of these states.
The conclusion and remarks about future work are contained in
Section~\ref{Spin32Conclusion}.

\section{Spin 3/2 Baryons on the Lattice}
\label{spin32sec}

\subsection{Spin 3/2 Interpolating Fields and Two-Point Functions}

In this Section the essential elements for a lattice calculation of
spin-$\frac{3}{2}$ baryon properties are presented.
The mass of a spin-${3\over 2}$ baryon on the lattice is obtained from
the two-point correlation function $G_{\mu\nu}$ \cite{LDW},
\be
G_{\mu\nu}(t, \vec p; \Gamma) = {\rm tr}_{\rm sp} \left\{ \Gamma {\cal
    G}_{\mu\nu}(t, \vec p) \right \}\, ,
\label{2ptfunc}
\ee
where
\begin{equation}
{\cal G}_{\mu\nu}^{\alpha\beta}(t, \vec p)
= \sum_{\vec x} e^{-i \vec p \cdot \vec x }\ 
  \langle 0 |\ T \left( \chi_\mu^\alpha (x)\ \overline \chi_\nu^\beta
    (0)  \right)
  | 0 \rangle\ ,
\label{CFunc}
\end{equation}
where $\chi_\mu^\alpha$ is a spin-${3\over 2}$ interpolating field,
$\Gamma$ is a matrix in Dirac space with $\alpha, \beta$ Dirac indices,
and $\mu, \nu$ Lorentz indices.

In this analysis we consider the following interpolating field
operator for the
isospin-${1\over 2}$, spin-${3\over 2}$, positive parity (charge $+1$)
state \cite{IOFFE} 
%
\be
\chi^N_{\mu} = \epsilon^{abc} 
\left( u^{Ta}(x)\ C \gamma_5 \gamma^\nu\ d^b(x) \right)
\left( g_{\mu\nu} - {1 \over 4} \gamma_\mu \gamma_\nu \right)
\gamma_5 u^c(x)\ .
\label{N32IFfull}
\ee
All discussions of interpolating fields are carried out using the
Dirac representation of the $\gamma$ matricies.
This exact isospin-${1\over 2}$ interpolating field has overlap with
both spin-${3\over 2}$ and spin-${1\over 2}$ states and with states of
both parities. The resulting correlation function will thus require
both spin and parity projection.
The quark field operators $u$ and $d$ act at Euclidean space-time point
$x$, $C$ is the charge conjugation matrix, $a, b$ and $c$ are colour
labels, and the superscript $T$ denotes the transpose.
The charge neutral interpolating field is obtained by interchanging
$u \leftrightarrow d$.
This interpolating field transforms as a Rarita-Schwinger operator under
parity transformations. That is, if the quark field operators transform as
$$
{\cal P} u(x) {\cal P}^\dagger = +\gamma_0 u(\tilde{x})\, ,
$$
where $\tilde{x} = (x_0, -\vec{x})$, and similarly for $d(x)$, then
$$
{\cal P} \chi^{N}_\mu(x) {\cal P}^\dagger
 = +\gamma_0 \chi^{N}_\mu(\tilde{x})\ ,
$$
and similarly for the Rarita-Schwinger operator
\be
{\cal P} u_\mu (x) {\cal P}^\dagger = +\gamma_0 u_\mu (\tilde{x})\ .
\ee
which will be discussed later.

The computational cost of evaluating each of the Lorentz combinations in
Eq.~(\ref{N32IFfull}) is relatively high --- about 100 times that for the
ground state nucleon \cite{LL}.
Consequently, in order to maximize statistics in our analysis we consider
only the leading term proportional to $g_{\mu\nu}$ in the
interpolating field,
\be
\chi^N_{\mu} \longrightarrow \epsilon^{abc}
\left( u^{Ta}(x)\ C \gamma_5 \gamma_\mu\ d^b(x) \right) \gamma_5 u^c(x)\ .
\label{N32IF}
\ee
This is sufficient since we will in either case need to perform a
spin-$\frac{3}{2}$ projection.

In order to show that the interpolating field defined in
Eq.~(\ref{N32IF}) has isospin-$\frac{1}{2}$, we first consider the standard
proton interpolating field, 
\be
\chi^p = \epsilon^{abc}(u^{Ta} C\gamma_5 d^b ) u^c,
\label{standardIF}
\ee  
which we know to have
isospin-$\frac{1}{2}$. Applying the isospin raising operator, $I^+$,
on $\chi^p$ one finds,
\begin{eqnarray*}
I^+ \chi^p &=& \epsilon^{abc}(u^{Ta} C\gamma_5 u^b ) u^c \\
&=& \epsilon^{abc}(u^{Ta} C\gamma_5 u^b )^T u^c \\
&=& -\epsilon^{abc}(u^{Ta} C\gamma_5 u^b ) u^c \\
&=& 0\, .
\end{eqnarray*}
Similarly, for the interpolating field defined in Eq.~(\ref{N32IF}),
one has
\begin{eqnarray*}
I^+ \chi_{\mu}^N &=& \epsilon^{abc}(u^{Ta} C\gamma_5 \gamma_{\mu} u^b )
\gamma_5 u^c \\
&=& \epsilon^{abc}(u^{Ta} C\gamma_5 \gamma_{\mu} u^b )^T \gamma_5 u^c \\
&=& -\epsilon^{abc}(u^{Ta} C\gamma_5 \gamma_{\mu} u^b ) \gamma_5 u^c \\
&=& 0\, ,
\end{eqnarray*}
where we have used the representation independent identities 
$C\gamma_{\mu}C^{-1} = -\gamma_{\mu}^T$, $C\gamma_5
C^{-1} = \gamma_5^T$ and the identities which hold in the Dirac
representation $C^T = C^\dagger = C^{-1} = -C$ with $C =
i\gamma_2\gamma_0$ and $\gamma_5^T = \gamma_5$.

We note that $\bar{\chi}_\mu^N$ corresponding to $\chi_\mu^N$ in
Eq.~(\ref{N32IF}) is
\be
\bar{\chi}_\mu^N = \epsilon^{abc} \bar{u}^a \gamma_{ 5}(\bar{d}^b
\gamma_{\mu}\gamma_{ 5} C \bar{u}^{cT} )\, ,
\ee
so that
\begin{eqnarray}
\chi_{\mu}^N \bar{\chi}_{\nu}^N \ =& \epsilon^{abc} \epsilon^{a'b'c'}
(u^{Ta}_{\alpha}[C\gamma_{5}\gamma_{\mu}]_{\alpha\beta}
d^b_{\beta})\gamma_{5}u^c_{\gamma} \bar{u}_{\gamma '}^{c'}\gamma_{5}
(\bar{d}^{b '}_{\beta '}[\gamma_{\nu}\gamma_{5} C]_{\beta ' \alpha '}
\bar{u}^{T a '}_{\alpha '})\ , \nonumber \\
%
\rightarrow& \gamma_5 S_u \gamma_5 {\rm tr} \left[\gamma_5 S_u \gamma_5
\left(C\gamma_{\mu} S_d \gamma_{\nu} C\right)^T \right] + \gamma_5 S_u \gamma_5
\left(C\gamma_{\mu} S_d \gamma_{\nu} C\right)^T \gamma_5 S_u \gamma_5\ ,
\end{eqnarray}
where the last line is the result achieved after doing the Grassman
integration over the quark fields with the quark fields being replaced
by all possible pairwise contractions.

In deriving the $\Delta$ interpolating fields, it is simplest to begin
with the state containing only valence $u$ quarks, namely the
$\Delta^{++}$.
The interpolating field for the $\Delta^{++}$ resonance
is given by \cite{IOFFE},
\begin{equation}
\chi_\mu^{\Delta^{++}}(x)
= \epsilon^{abc}
\left( u^{Ta}(x)\ C \gamma_\mu\ u^b(x) \right) u^c(x)\ ,
\label{deltaIF}
\end{equation}
which also transforms as a pseudovector under parity.
The interpolating field for a $\Delta^+$ state can be similarly
constructed \cite{LDW},
\begin{eqnarray}
\chi_\mu^{\Delta^{+}}(x)
= {1 \over \sqrt{3} } \; \epsilon^{abc}
\left[ 2 \left( u^{Ta}(x)\ C \gamma_\mu\ d^b(x) \right)\ u^c(x)\
      +\ \left( u^{Ta}(x)\ C \gamma_\mu\ u^b(x) \right)\ d^c(x)
\right]\ .
\end{eqnarray}
Interpolating fields for other decuplet baryons are obtained by
appropriate substitutions of $u,\ d\ \to\ u,\ d$ or $s$
fields.

To project a pure spin-${3\over 2}$ state from the correlation function
$G_{\mu\nu}$, one needs to use an appropriate spin-${3\over 2}$ projection
operator \cite{BDM},
\begin{equation}
P^{3/2}_{\mu \nu}(p)
= g_{\mu \nu}
- {1 \over 3} \gamma_\mu \gamma_\nu
- {1 \over 3 p^2}
   \left( \gamma \cdot p\, \gamma_\mu\, p_\nu
        + p_\mu\, \gamma_\nu\, \gamma \cdot p
   \right)\ .
\end{equation}
The corresponding spin-${1\over 2}$ state can be projected by applying the
projection operator
\begin{equation}
P_{\mu\nu}^{1/2} = g_{\mu\nu} - P_{\mu\nu}^{3/2}\ .
\end{equation}
To use this operator and retain all Lorentz components,
one must calculate the full $4\times 4$ matrix in Dirac and Lorentz space.
However, to extract a mass, only one pair of Lorentz indices is needed, reducing
the amount of calculations required by a factor of four. We calculate the third
row of the Lorentz matrix and use the projection,
\be
G^s_{33} = \sum_{\mu,\nu = 1}^4 G_{3\mu}\, g^{\mu\nu}\, P_{\nu 3}^s\ ,
\label{SpinPrCF}
\ee
to extract the desired spin states, $s=\frac{1}{2}$ or $\frac{3}{2}$.
Following spin projection, the resulting correlation function, $G^s_{33}$,
still contains positive and negative parity states.

\subsection{Baryon Level}

The interpolating field operators defined in Eqs.~(\ref{N32IFfull}) and (\ref{N32IF})
have overlap with both spin-$\frac{3}{2}$ and spin-$\frac{1}{2}$ states with
positive and negative parity.
The field $\chi_\mu$ transforms as a pseudovector under parity, as does
the Rarita-Schwinger spinor, $u_\mu$. Thus the overlap of $\chi_\mu$
with baryons can be expressed as
\begin{subequations}
\begin{eqnarray}
\langle 0 | \chi_{\mu} | N^{\frac{3}{2}+}(p,s) \rangle &=&
\lambda_{3/2^+} \sqrt{ {M_{3/2^+} \over E_{3/2^+}} }\ u_{\mu}(p,s)\ , \\
\langle 0 | \chi_{\mu} | N^{\frac{3}{2}-}(p,s) \rangle &=&
\lambda_{3/2^-} \sqrt{ {M_{3/2^-} \over E_{3/2^-}} }\ \gamma_5
u_{\mu}(p,s)\ , \\
\langle 0 | \chi_{\mu} | N^{\frac{1}{2}+}(p,s) \rangle &=&
(\alpha_{1/2^+}p_{\mu} + \beta_{1/2^+}\gamma_{\mu})
\sqrt{ {M_{1/2^+} \over E_{1/2^+}} }\ \gamma_5 u(p,s)\ , \label{third}\\
\langle 0 | \chi_{\mu} | N^{\frac{1}{2}-}(p,s) \rangle &=&
(\alpha_{1/2^-}p_{\mu} + \beta_{1/2^-}\gamma_{\mu})
\sqrt{ {M_{1/2^-} \over E_{1/2^-}} }\ u(p,s)\ , \label{fourth}
\end{eqnarray}
\end{subequations}
%
where the factors $\lambda_B ,\, \alpha_B ,\,
\beta_B$ denote the coupling strengths of the interpolating field
$\chi_\mu$ to the baryon $B$ and $E_B = \sqrt{\vec p\,^2 + M_B^2}$ is the
energy.  For the expressions in Eqs.~(\ref{third}) and (\ref{fourth}),
we note that the 
spatial components of momentum, $p_i$, transform as a vector under
parity and commute with $\gamma_0$, whereas the $\gamma_i$ do not
change sign under parity but anticommute with $\gamma_0$. Hence the
right-hand-side of Eq.~(\ref{third}) also transforms as a
pseudovector under parity in accord with $\chi_\mu$.

Similar expressions can also be written for $\bar{\chi}_\mu$,
\begin{subequations}
\begin{eqnarray}
\langle N^{\frac{3}{2}+}(p,s) | \bar{\chi}_{\mu} | 0 \rangle &=&
\lambda_{3/2^+}^* \sqrt{ {M_{3/2^+} \over E_{3/2^+}} }\
\bar{u}_{\mu}(p,s)\ ,\\
\langle N^{\frac{3}{2}-}(p,s) | \bar{\chi}_{\mu} | 0 \rangle 
&=& -\lambda_{3/2^-}^* \sqrt{ {M_{3/2^-} \over E_{3/2^-}} }\
\bar{u}_{\mu}(p,s) \gamma_5\ ,\\
\langle N^{\frac{1}{2}+}(p,s) | \bar{\chi}_{\mu} | 0 \rangle 
%
&=& - \sqrt{ {M_{1/2^+} \over E_{1/2^+}} }\ \bar{u}(p,s) \gamma_5
(\alpha^*_{1/2^+}p_{\mu} + \beta^*_{1/2^+}
\gamma_{\mu})\ , \\
\langle N^{\frac{1}{2}-}(p,s) | \bar{\chi}_{\mu} | 0 \rangle 
%
&=& \sqrt{ {M_{1/2^-} \over E_{1/2^-}} }\ \bar{u}(p,s)
(\alpha^*_{1/2^-}p_{\mu} + \beta^*_{1/2^-}
\gamma_{\mu})\ .
\end{eqnarray}
\end{subequations}
Note that we are assuming identical sinks and sources in these
equations. In our calculations we use a smeared source and a point
sink in which case $\lambda^* ,\ \alpha^*$ and $\beta^*$ are no longer
complex conjugates of $\lambda,\ \alpha$ and $\beta$ and are instead
replaced by $\overline{\lambda},\ \overline{\alpha}$ and
$\overline{\beta}$. 

We are now in a position to find the form of Eq.~(\ref{CFunc}) after we
insert a complete set of intermediate states $\left\{| B (p,s)
  \rangle\right\}$. 
The contribution to Eq.~(\ref{CFunc}) from each intermediate state
considered is given by 
\begin{eqnarray*}
\langle 0 | \chi_{\mu} | N^{\frac{3}{2}+}(p,s) \rangle 
\!\!\!\!\! && \!\!\!\!\! 
\langle N^{\frac{3}{2}+}(p,s) | \bar{\chi}_{\nu} | 0 \rangle \\
&=& + \lambda_{3/2^+}\overline{\lambda}_{3/2^+} \,{M_{3/2^+} \over
  E_{3/2^+}} u_\mu (p,s) \bar{u}_\nu (p,s) \\
&=& - \lambda_{3/2^+}\overline{\lambda}_{3/2^+} \, {M_{3/2^+} \over E_{3/2^+}} 
{ (\gamma \cdot p + M_{3/2^+}) \over 2 M_{3/2^+} }
 \left\{ g_{\mu\nu}
        - { 1 \over 3} \gamma_\mu \gamma_\nu
        - { 2 p_\mu p_\nu \over 3 M^2_{3/2^+} }
        + { p_\mu \gamma_\nu - p_\nu \gamma_\mu \over 3 M_{3/2^+}}
  \right\}\ , \\
\langle 0 | \chi_{\mu} | N^{\frac{3}{2}-}(p,s) \rangle 
\!\!\!\!\! && \!\!\!\!\! 
\langle N^{\frac{3}{2}-}(p,s) | \bar{\chi}_{\nu} | 0 \rangle \\
&=& 
- \lambda_{3/2^-}\overline{\lambda}_{3/2^-} \, {M_{3/2^-} \over
  E_{3/2^-}} \gamma_5 u_\mu (p,s) \bar{u}_\nu (p,s)
\gamma_5 \\
%
&=& -\lambda_{3/2^-}\overline{\lambda}_{3/2^-} \, {M_{3/2^-} \over E_{3/2^-}} 
{ (\gamma \cdot p - M_{3/2^-}) \over 2 M_{3/2^-} }
  \left\{ g_{\mu\nu}
        - { 1 \over 3} \gamma_\mu \gamma_\nu
        - { 2 p_\mu p_\nu \over 3 M^2_{3/2^-} }
        - { p_\mu \gamma_\nu - p_\nu \gamma_\mu \over 3 M_{3/2^-}}
  \right\}\ , \\
\langle 0 | \chi_{\mu} | N^{\frac{1}{2}+}(p,s) \rangle 
\!\!\!\!\! && \!\!\!\!\! 
\langle N^{\frac{1}{2}+}(p,s) | \bar{\chi}_{\nu} | 0 \rangle \\
&=& -{M_{1/2^+} \over E_{1/2^+}}
(\alpha_{1/2^+}p_{\mu} + \beta_{1/2^+}
\gamma_{\mu}) \gamma_5 \frac{\gamma\cdot p + M_{1/2^+}}{2M_{1/2^+}} \gamma_5
(\overline\alpha_{1/2^+}p_{\nu} + \overline\beta_{1/2^+}
\gamma_{\nu})\ , \\
\langle 0 | \chi_{\mu} | N^{\frac{1}{2}-}(p,s) \rangle 
\!\!\!\!\! && \!\!\!\!\! 
\langle N^{\frac{1}{2}-}(p,s) | \bar{\chi}_{\nu} | 0 \rangle \\
&=& {M_{1/2^-} \over E_{-/2^+}}
(\alpha_{1/2^-}p_{\mu} + \beta_{1/2^-}
\gamma_{\mu}) \frac{\gamma\cdot p + M_{1/2^-}}{2M_{1/2^-}} 
(\overline\alpha_{1/2^-}p_{\nu} + \overline\beta_{1/2^-}
\gamma_{\nu})\ .
\end{eqnarray*}
To reduce computational expense, we consider the specific case when
$\mu = \nu = 3$ and in order to extract masses we require $\vec{p} =
(0,0,0)$. In this case we have the simple expressions
\begin{subequations}
\begin{eqnarray}
\langle 0 | \chi_{3} | N^{\frac{3}{2}+}(p,s) \rangle 
\langle N^{\frac{3}{2}+}(p,s) | \bar{\chi}_{3} | 0 \rangle
&=& 
\lambda_{3/2^+}\overline{\lambda}_{3/2^+} \frac{2}{3}
\left( { \gamma_0 M_{3/2^+} + M_{3/2^+} \over 2 M_{3/2^+} } \right)\, , \\
\langle 0 | \chi_{3} | N^{\frac{3}{2}-}(p,s) \rangle 
\langle N^{\frac{3}{2}-}(p,s) | \bar{\chi}_{3} | 0 \rangle
&=& 
\lambda_{3/2^-}\overline{\lambda}_{3/2^-} \frac{2}{3}
\left( { \gamma_0 M_{3/2^-} - M_{3/2^-} \over 2 M_{3/2^-} } \right)\, , \\
\langle 0 | \chi_{3} | N^{\frac{1}{2}+}(p,s) \rangle 
\langle N^{\frac{1}{2}+}(p,s) | \bar{\chi}_{3} | 0 \rangle 
&=& - \beta_{1/2^+}\overline{\beta}_{1/2^+} \gamma_3 \gamma_5 \frac{\gamma_0 M_{1/2^+} +
  M_{1/2^+}}{2M_{1/2^+}} \gamma_5 \gamma_3 \nonumber \\
&=& + \beta_{1/2^+}\overline{\beta}_{1/2^+} \frac{\gamma_0 M_{1/2^+} +
  M_{1/2^+}}{2M_{1/2^+}}\, , \\
\langle 0 | \chi_{3} | N^{\frac{1}{2}-}(p,s) \rangle 
\langle N^{\frac{1}{2}-}(p,s) | \bar{\chi}_{3} | 0 \rangle 
&=& \beta_{1/2^-}\overline{\beta}_{1/2^-} \gamma_3 \frac{\gamma_0 M_{1/2^-} +
  M_{1/2^-}}{2M_{1/2^-}} \gamma_3 \nonumber \\
&=& + \beta_{1/2^-}\overline{\beta}_{1/2^-} \frac{\gamma_0 M_{1/2^-} -
  M_{1/2^-}}{2M_{1/2^-}}\, .
\end{eqnarray}
\end{subequations}
Therefore, in an analogous procedure to that used in
Ref.~\cite{NSTAR}, where a fixed boundary condition is used in the
time direction, positive and negative parity states are obtained by
taking the trace of the spin-projected correlation function,
$G^s_{33}$, in Eq.~(\ref{SpinPrCF}) with the operator 
$\Gamma = \Gamma_{\pm}$, 
\be
G^{s\pm}_{33} = {\rm tr_{sp}}\left\{\Gamma^{\pm}\, G^s_{33}\right\}\, ,
\ee
where
\be
\Gamma_{\pm} = {1\over 2}\left( 1\pm \gamma_4 \right) .
\ee
The positive parity states propagate in the (1,1) and (2,2)
elements of the Dirac matrix, while negative parity states propagate
in the (3,3) and (4,4) elements for both spin-$\frac{1}{2}$ and
spin-$\frac{3}{2}$ projected states. A similar treatment has been
carried out for the $\Delta$ interpolating fields but is not
reproduced here for brevity.

We use an improved unbiased estimator obtained by summing both $U$ and
$U^*$ configurations which occur with equal weight. 
From the discussion given in Section~Va of Ref~\cite{NSTAR},
$G_{\mu\nu}^{s\pm}$ is purely real if $\mu$ and $\nu$ are both
spatial indicies or both temporal indicies, otherwise
$G_{\mu\nu}^{s\pm}$ is purely imaginary. 


\section{Results}
\label{spin32results}

The analysis is based on a sample of 392 configurations.
For the gauge fields, a mean-field improved plaquette plus rectangle
action is used. 
The simulations are performed on a $16^3\times 32$ lattice at $\beta=4.60$,
which corresponds to a lattice spacing of $a = 0.122(2)$~fm set by a
string tension analysis incorporating the lattice coulomb potential
\cite{Edwards:1998xf} with $\sqrt\sigma = 440$~MeV.
For the quark fields, a Fat-Link Irrelevant Clover
(FLIC) \cite{FATJAMES} action is implemented.
The use of fat links \cite{FATLINK} in the irrelevant operators of the
fermion action removes the need to fine tune the clover coefficient to
remove ${\cal O}(a)$ artifacts.
The use of fat links also allows us to employ a highly improved definition of
$F_{\mu\nu}$ \cite{FATJAMES,SUNDANCE} leaving errors of ${\cal
  O}(a^6)$ and where errors of ${\cal O}(g^2)$ are suppressed via fat links.
Mean-field improvement of the tree-level clover coefficient with fat
links represents a small correction and proves to be adequate
\cite{FATJAMES}.

The fattening, or smearing of the lattice links with their nearest
neighbours, reduces the problem of exceptional configurations, and
minimizes the effect of renormalization on the action improvement terms.
By smearing only the irrelevant, higher dimensional terms in the action,
and leaving the relevant dimension-four operators untouched, we retain
short distance quark and gluon interactions at the scale of the cutoff.
Simulations are performed with $n=4$ smearing sweeps and a smearing
fraction $\alpha=0.7$ \cite{FATJAMES}.

A fixed boundary condition in the time direction is used for the fermions
by setting $U_t(\vec x, N_t) = 0\ \forall\ \vec x$ in the hopping terms
of the fermion action, with periodic boundary conditions imposed in the
spatial directions.
Gauge-invariant gaussian smearing \cite{Gusken:qx} in the spatial
dimensions is applied at the source to increase the overlap of the
interpolating operators with the ground states.
The source-smearing technique \cite{Gusken:qx} starts with a point
source, 
\be
\psi_{0\, \alpha}^{\phantom{0}\, a}({\vec x}, t) = \delta^{ac}
\delta_{\alpha\gamma} \delta_{{\vec x},{\vec x}_0} \delta_{t,t_0}
\label{ptsource}
\ee
for source colour $c$, Dirac $\gamma$, position ${\vec x}_0 = (1,1,1)$
and time $t_0 = 3$ and proceeds via the iterative scheme, 
\[
\psi_i({\vec x},t) = \sum_{{\vec x}'} F({\vec x},{\vec x}') \, \psi_{i-1}({\vec x}',t) \, ,
\]
where
\[
F({\vec x},{\vec x}') = \frac{1}{(1+\alpha)} \left( \delta_{{\vec x},
    {\vec x}'} + 
  \frac{\alpha}{6} \sum_{\mu=1}^3 \left [ U_\mu({\vec x},t) \, \delta_{{\vec x}',
{\vec x}+\widehat\mu} + 
U_\mu^\dagger({\vec x}-\widehat\mu,t) \, \delta_{{\vec x}', {\vec
    x}-\widehat\mu} \right ] \right) \, . 
\]
Repeating the procedure $N$ times gives the resulting fermion source
\be
\psi_N({\vec x},t) = \sum_{{\vec x}'} F^N({\vec x},{\vec x}') \, \psi_0({\vec x}',t) \, .
\ee
The parameters $N$ and $\alpha$ govern the size and shape of the
smearing function. We simulate with $N=20$ and $\alpha=6$.
The propagator, $S$, is obtained from the smeared source by
solving 
\be
M_{\alpha\beta}^{ab}\, S_{\beta\gamma}^{bc} =
\psi_{\alpha}^{a}\, ,
\ee
for each colour, Dirac source $c,\, \gamma$ respectively of
Eq.~(\ref{ptsource}) via the BiStabilised Conjugate Gradient algorithm
\cite{BiCG}.

\begin{table}[t]
\begin{center}
\caption{Masses of the $\pi,\, N{\frac{1}{2}^\pm}$ and
  $N{\frac{3}{2}^\pm}$, for several values of $\kappa$ obtained from
  the spin-$\frac{3}{2}$ interpolating field, for the FLIC
  action with 4 sweeps of smearing at $\alpha=0.7$. Here the value of
  $\kappa_{\rm cr}$ is $\kappa_{\rm cr} = 0.1300$.  A string tension
  analysis provides $a=0.122(2)$ fm for
  $\sqrt\sigma=440$~MeV.
\label{kappaN}} 
\vspace*{0.5cm}
\begin{ruledtabular}
  \begin{tabular}{cccccc}
        $\kappa$ & $m_{\pi}\, a$ &
        $M_{N{\frac{1}{2}+}}\, a$ & $M_{N{\frac{1}{2}-}}\, a$ &
        $M_{N{\frac{3}{2}+}}\, a$ & $M_{N{\frac{3}{2}-}}\, a$   \\ \hline
    0.1260 &  0.5767(11)  &  1.102(8)  &   1.412(13) &  1.628(34)  &  1.410(16) \\
    0.1266 &  0.5305(12)  &  1.043(9)  &   1.369(14) &  1.577(38)  &  1.365(19) \\
    0.1273 &  0.4712(15)  &  0.970(13)  &  1.317(17) &  1.510(44)  &  1.312(24) \\
    0.1279 &  0.4164(15)  &  0.905(18)  &  1.271(21) &  1.440(53)  &  1.264(32) \\
    0.1286 &  0.3421(18)  &  0.829(32)  &  1.220(31) &  1.329(74)  &  1.206(49) 
\end{tabular}
\end{ruledtabular}
\end{center}
\end{table}

\begin{figure}
\begin{center}
\includegraphics[height=0.75\hsize,angle=90]{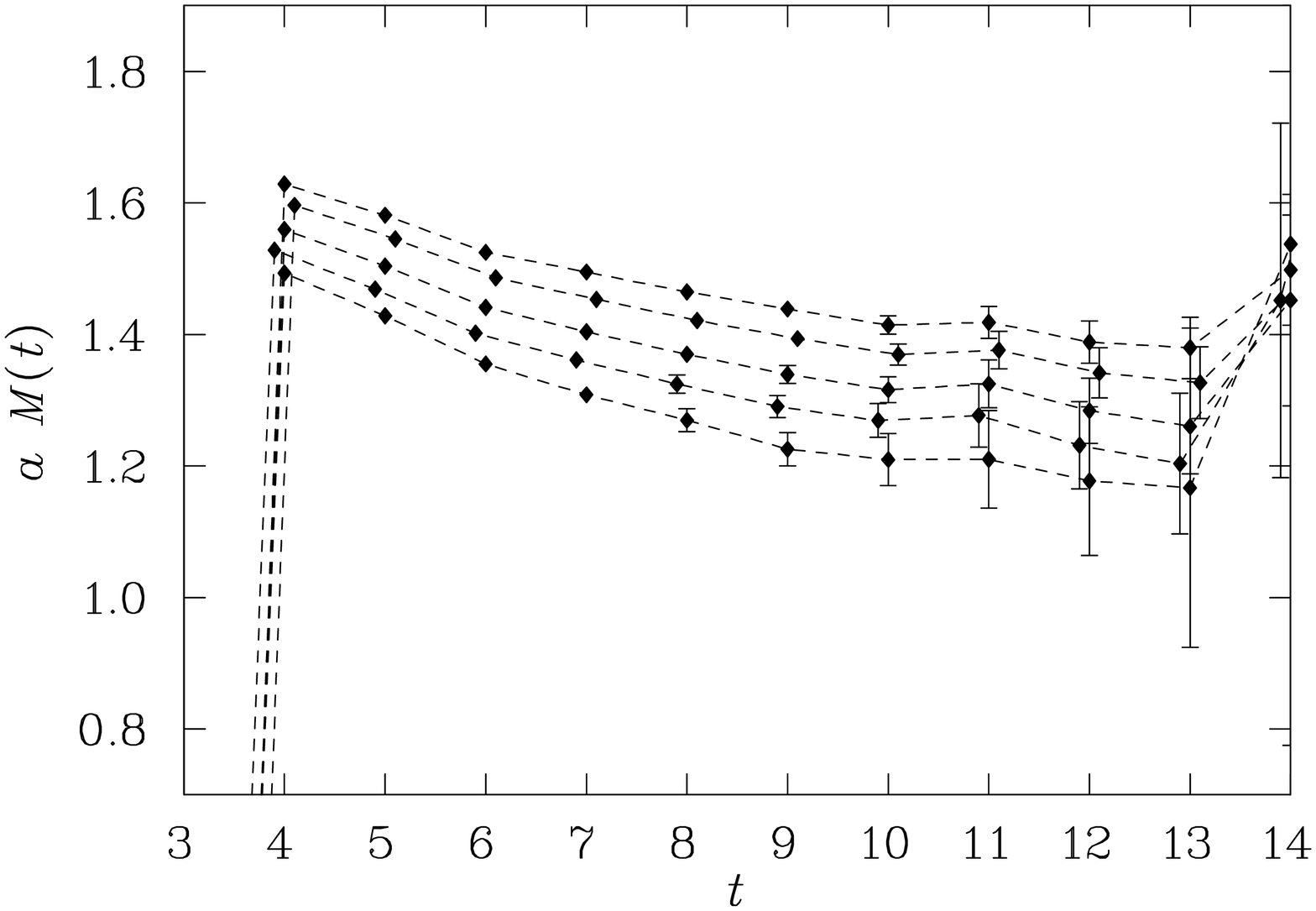} 
\vspace*{0.35cm}
\caption{Effective mass plot for the $N{3\over 2}^-$ state
  using the FLIC action, from 392 configurations. The five sets of
  points correspond to the $\kappa$ values listed in
  Table~\protect\ref{kappaN}, with $\kappa$ increasing from top down.
  \label{N32-Mass}}
\vspace*{1.3cm}
\includegraphics[height=0.75\hsize,angle=90]{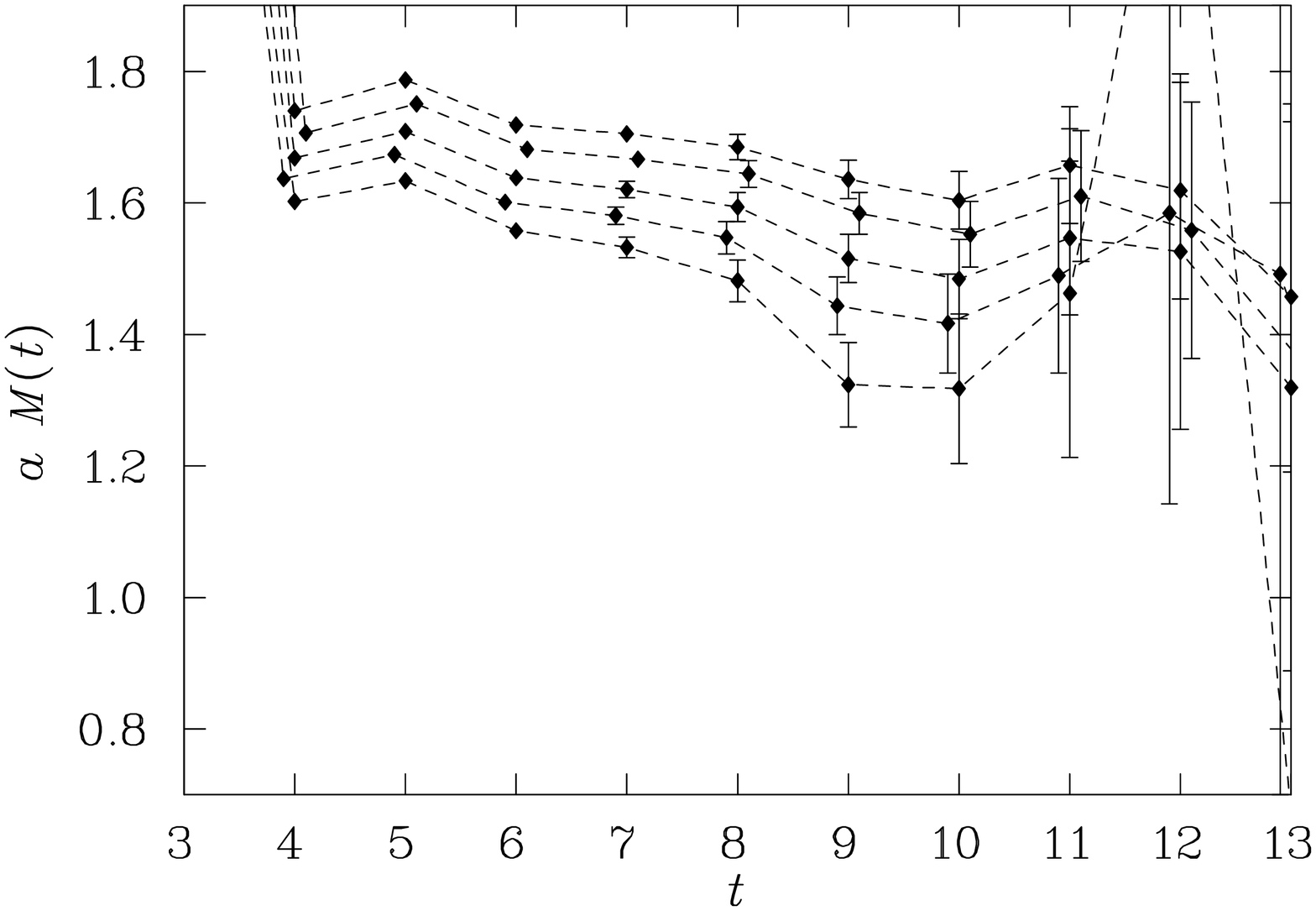} 
\vspace*{0.35cm}
\caption{As in Fig.~\ref{N32-Mass}, but for the $N{3\over 2}^+$ state.
  \label{N32+Mass}}
\end{center}
\end{figure}

In the analysis we use five values of $\kappa$, as indicated in
Table~\ref{kappaN}. Extrapolation to $m_\pi^2 = 0$ gives $\kappa_{\rm
  cr} = 0.1300$.  
Figure~\ref{N32-Mass}
shows the effective mass plot for the $N{3\over 2}^-$ state for the
five $\kappa$ values used as a
function of Euclidean time obtained after performing spin and parity
projections on the correlation functions calculated using the
interpolating field in Eq.~(\ref{N32IF}). We find a good signal
for this state up until time slice 13 after which the signal is lost
in noise.
The effective mass for this state exhibits good plateau behaviour and
a good value of the covariance-matrix based $\chi^2 / N_{\rm DF}$ is 
obtained when one fits in the time fitting window of $t=10$--13
(recall, the source is at $t=3$).
Typically, one finds $\chi^2 / N_{\rm DF} \approx 1$ and $\chi^2 /
N_{\rm DF} < 1.5$ throughout.
After performing spin and parity projections to extract the $N{3\over 2}^+$
state from the interpolating field in Eq.~(\ref{N32IF}), one finds
the effective mass plot to be a little noisier, as shown in
Fig.~\ref{N32+Mass}.
There is, however, sufficient information
here to extract a mass, and a good value of $\chi^2 / N_{\rm DF}$ is
obtained when one fits in the small time fitting window of $t=9$--11.

\begin{figure}
\begin{center}
\includegraphics[height=0.75\hsize,angle=90]{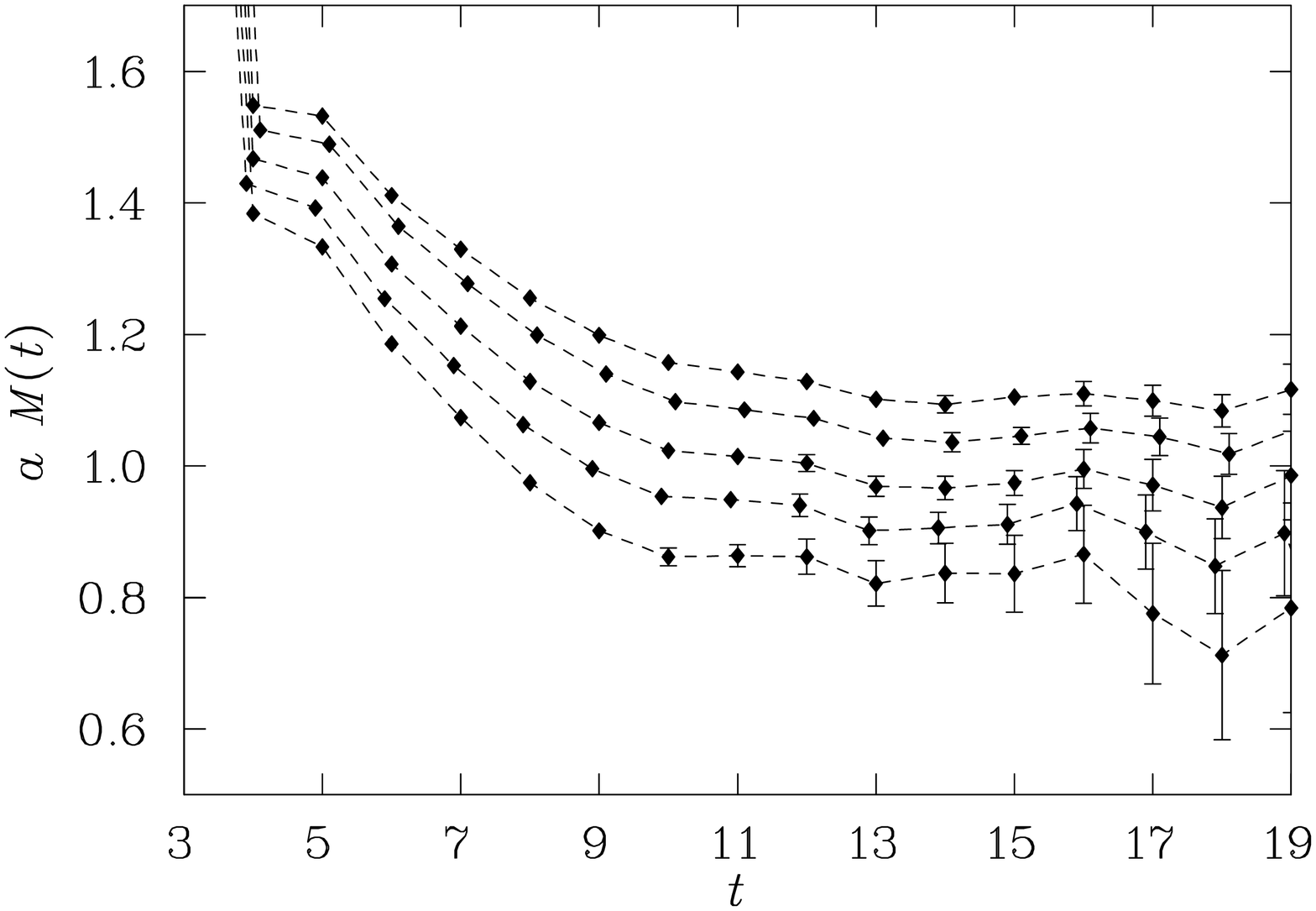} 
\vspace*{0.35cm}
\caption{As in Fig.~\ref{N32-Mass}, but for the $N{1\over 2}^+$ state.
\label{N12+Mass}}
\vspace*{1.3cm}
\includegraphics[height=0.75\hsize,angle=90]{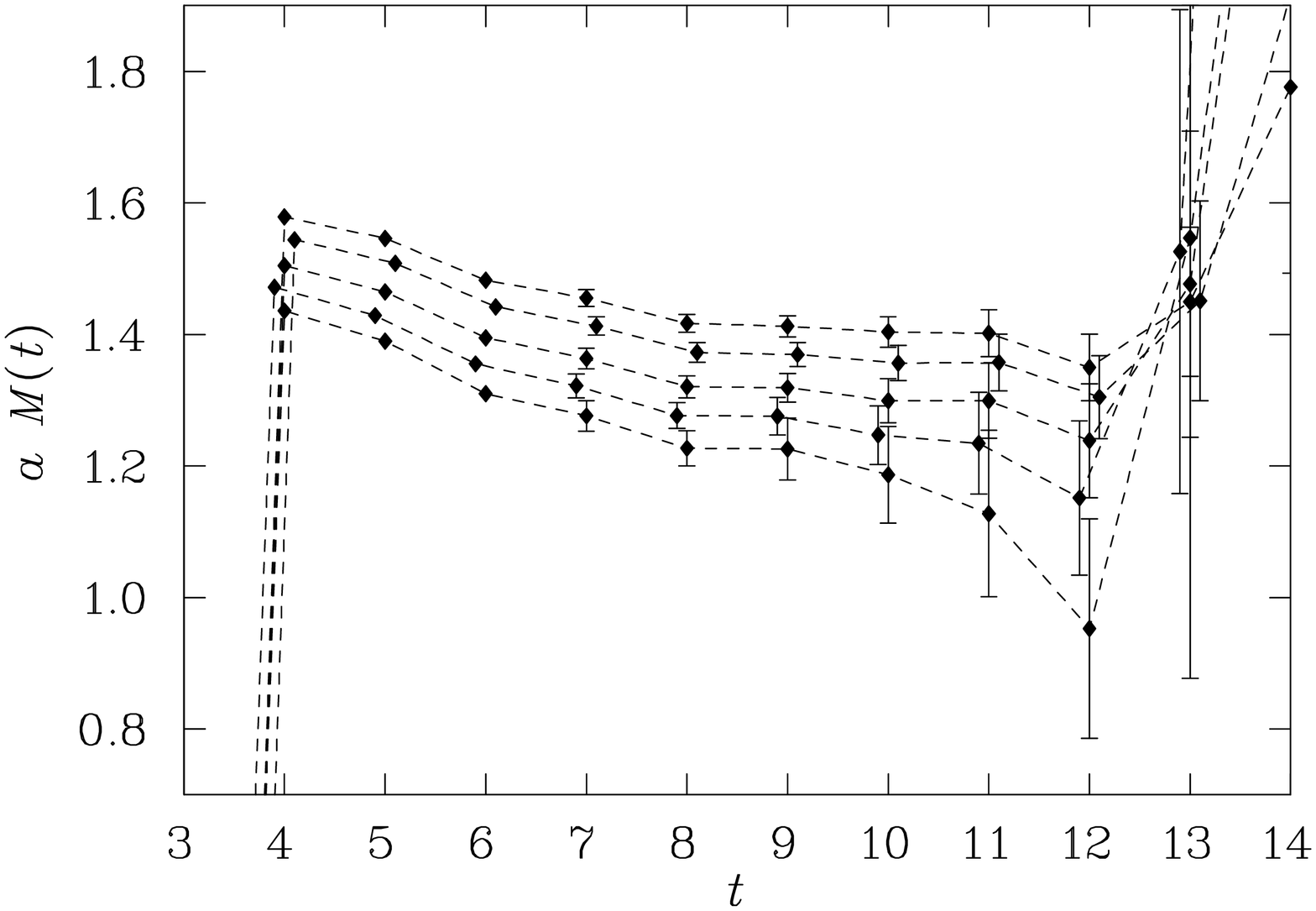} 
\vspace*{0.35cm}
\caption{As in Fig.~\ref{N32-Mass}, but for the $N{1\over 2}^-$ state.
\label{N12-Mass}}
\end{center}
\end{figure}

The interpolating field defined in Eq.~(\ref{N32IF}) also has overlap
with spin-$\frac{1}{2}$ states of both parities. After performing a
spin-$\frac{1}{2}$ projection on the correlation functions, we isolate
the $N{1\over 2}^+$ and $N{1\over 2}^-$ states via parity
projection and plot the effective masses in Figs.~\ref{N12+Mass} and
\ref{N12-Mass} respectively. The $N{1\over 2}^+$ state suffers
contamination from excited states as seen by the long Euclidean
time evolution required to reach plateau in Fig.~\ref{N12+Mass}. A
good value of $\chi^2 / N_{\rm DF}$ is obtained as long as we fit
after time slice 12. For this reason, we use time slices 13--16 to
obtain a mass for the $N{1\over 2}^+$ state. However, for the
$N{1\over 2}^-$ state, a plateau is seen at early Euclidean times
and a good value of $\chi^2 / N_{\rm DF}$ is obtained on time slices
8--11.

\begin{figure}[t]
\begin{center}
\includegraphics[height=0.85\hsize,angle=90]{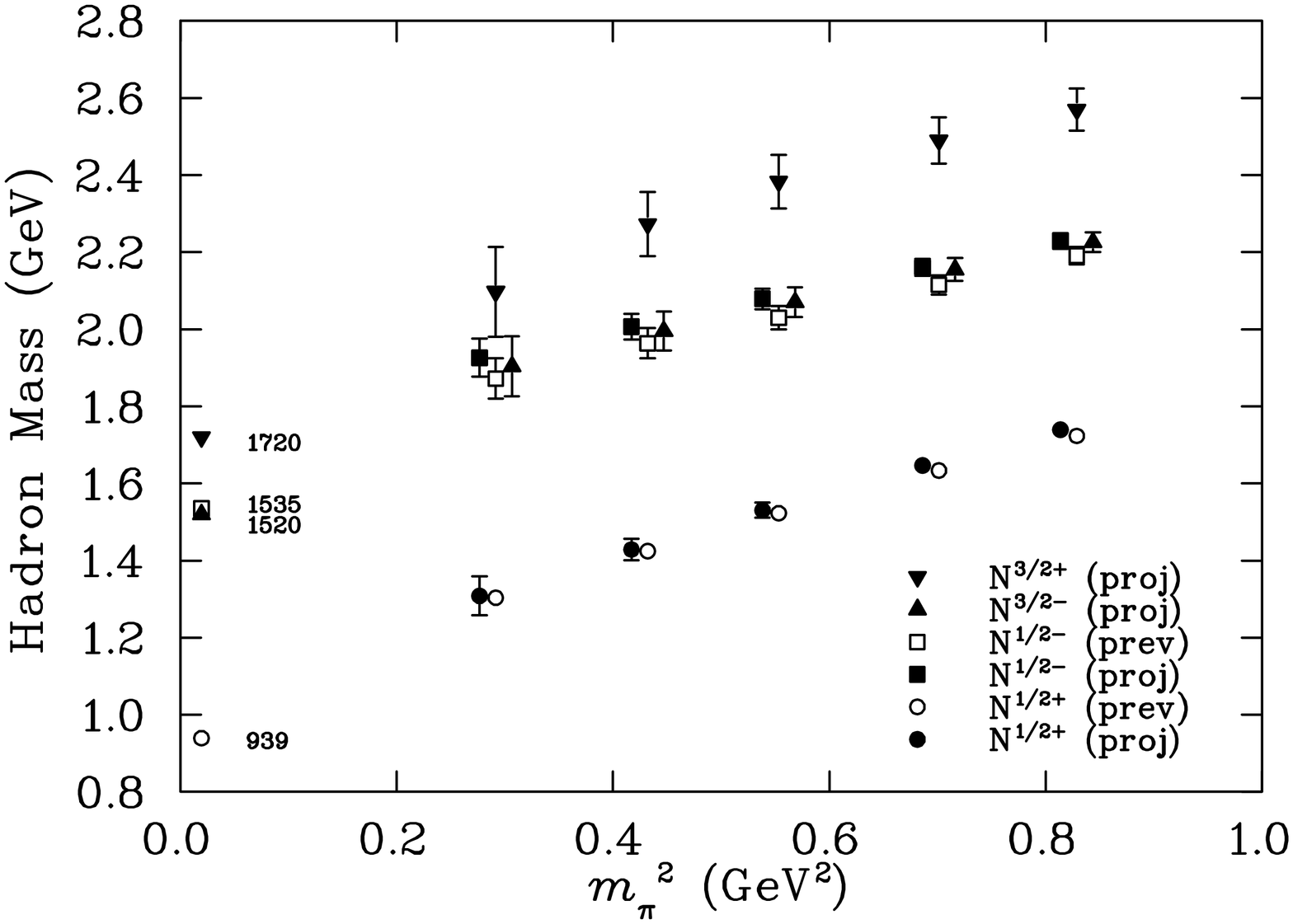} 
\vspace*{0.5cm}
\caption{Masses of the spin projected $N{3\over 2}^-$ (filled triangles),
  $N{3\over 2}^+$ (filled inverted triangles), $N{1\over 2}^+$
  (filled circles), and $N{1\over 2}^-$ (filled squares) states. For
  comparison, previous results from the direct calculation of the
  $N{1\over 2}^+$ (open circles) and $N{1\over 2}^-$ (open squares)
  from Ref.~\protect\cite{NSTAR} are also shown. The empirical values
  of the masses of the $N{1\over 2}^+\,(939)$, $N{1\over 2}^-\,
  (1535)$, $N{3\over 2}^-\, (1520)$ and $N{3\over 2}^+\, (1720)$  are
  shown on the left-hand-side at the physical pion mass.
\label{Nvsmpi2}}
\end{center}
\end{figure}

The extracted masses of the $N{3\over 2}^\pm$ and $N{1\over 2}^\pm$
states are given in Table~\ref{kappaN} and are displayed in
Fig.~\ref{Nvsmpi2} as a function of $m_\pi^2$. 
Earlier results for the $N\frac{1}{2}^\pm$ states using the standard
spin-$1\over 2$ interpolating field \cite{NSTAR,FATJAMES} from
Eq.~(\ref{standardIF}) are also shown with open symbols in
Fig.~\ref{Nvsmpi2} for reference.
It is encouraging to note the agreement between the
spin-projected ${1\over 2}^\pm$ states obtained from the spin-$3\over 2$
interpolating field in Eq.~(\ref{N32IF}) and the earlier ${1\over 2}^\pm$
results from the same gauge field configurations.
To study this agreement more accurately, we
consider the ratio of effective masses obtained for each
jackknife subensemble. This provides us with a correlated ratio and we
find the ratio to be one as at the one standard deviation level.
We also observe that the $N\frac{3}{2}^-$ state has approximately the
same mass as the spin-projected $N\frac{1}{2}^-$ state which is
consistent with the experimentally observed masses.
To study this mass difference more accurately, we
again calculate the correlated ratio of effective masses obtained after
appropriate spin and parity projections. 
This ratio is found to be one within one standard deviation.
The results for the $N\frac{3}{2}^-$ state in Fig.~\ref{Nvsmpi2}
indicate a clear mass splitting between the $N{3\over 2}^+$ and
$N{3\over 2}^-$ states obtained from the spin-${3\over 2}$
interpolating field, with a mass difference around 300~MeV. This is
slightly larger than the experimentally observed mass difference of
200~MeV.

\begin{figure}[t]
\begin{center}
\includegraphics[height=0.75\hsize,angle=90]{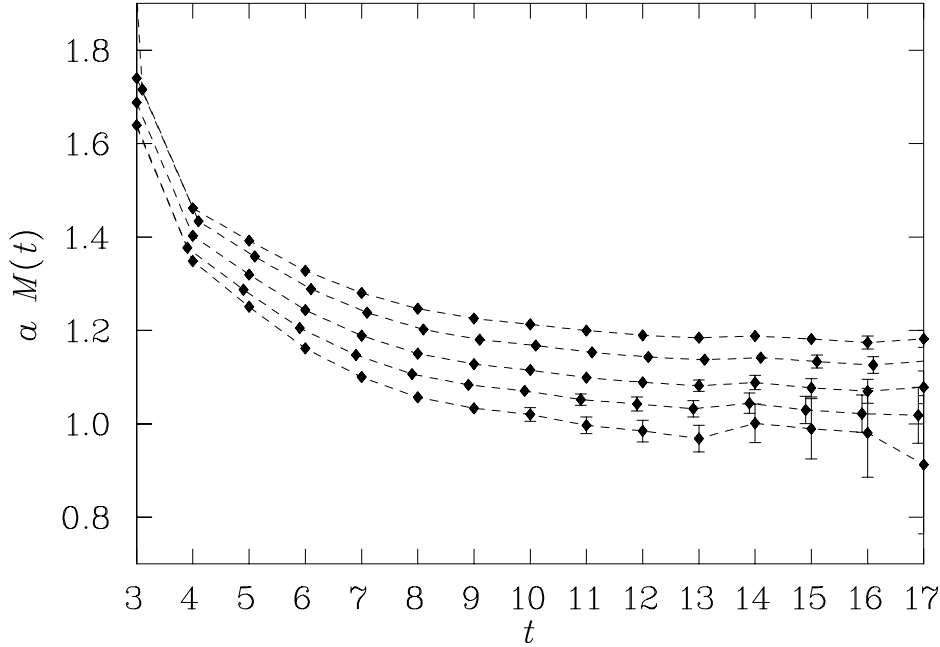} 
\vspace*{0.35cm}
\caption{Effective mass plot for the $\Delta{3\over 2}^+$ state using
        the FLIC action with 4 sweeps of smearing at $\alpha = 0.7$ from
        392 configurations.  The five sets of points correspond to the
        $\kappa$ values listed in Table~\protect\ref{kappaD}, with
        $\kappa$ increasing from top down.
\label{D32+Mass}}
\end{center}
\end{figure}

\begin{figure}[t]
\begin{center}
\includegraphics[height=0.75\hsize,angle=90]{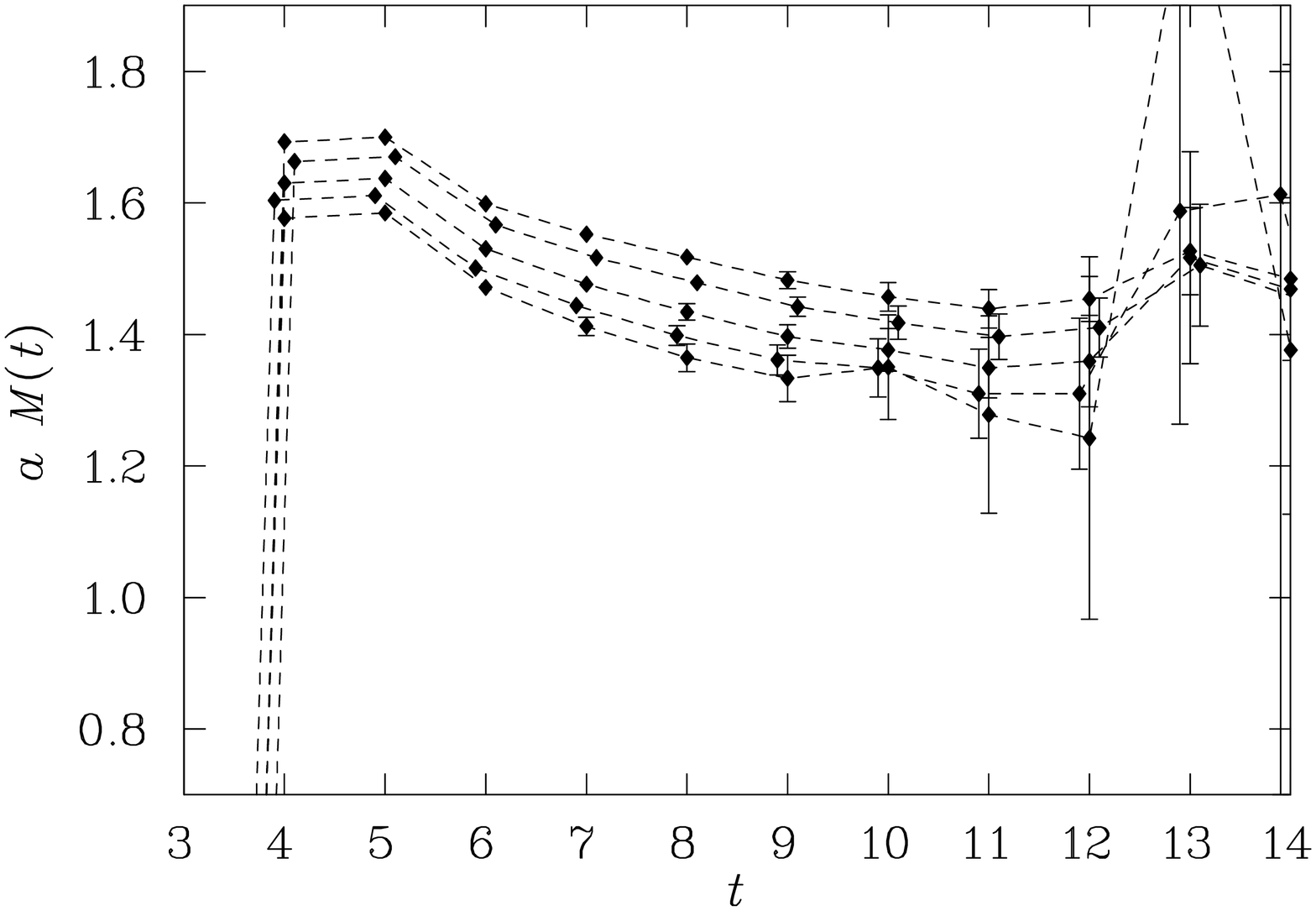} 
\vspace*{0.35cm}
\caption{As in Fig.~\ref{D32+Mass}, but for the $\Delta{3\over 2}^-$ state.
\label{D32-Mass}}
\end{center}
\end{figure}

Turning now to the isospin-${3\over 2}$ sector, the effective
mass plot for the $\Delta{3\over 2}^+$ 
state using the interpolating field given in Eq.~(\ref{deltaIF}) is
shown in Figure~\ref{D32+Mass} for the five $\kappa$ values used.  An excellent
signal is clearly visible, and a good value of the covariance-matrix
based $\chi^2 / N_{\rm DF}$ is obtained by fitting time slices
$t=11$--14 following the source at $t=3$.
For the effective mass of the negative parity $\Delta{3\over 2}^-$,
shown in Fig.~\ref{D32-Mass}, the signal is quite good up to time slice 11-12, but
is lost in noise after time slice 12.  Time slices $t=9$--12 provide a
fitting window with an acceptable value of $\chi^2 / N_{\rm DF}$.

\begin{table}
\begin{center}
\caption{As in Table~\ref{kappaN}, but for the corresponding
  $\Delta^{\frac{3}{2}+},\, \Delta^{\frac{3}{2}-},\,
  \Delta^{\frac{1}{2}-}$ and $\Delta^{\frac{1}{2}+}$ masses.
\label{kappaD}}
\vspace*{0.5cm}
\begin{ruledtabular}
  \begin{tabular}{ccccc}
        $\kappa$ &  $M_{\Delta^{\frac{3}{2}+}}\, a$ & 
        $M_{\Delta^{\frac{3}{2}-}}\, a$ & 
        $M_{\Delta^{\frac{1}{2}+}}\, a$ & 
        $M_{\Delta^{\frac{1}{2}-}}\, a$  \\ \hline
0.1260 &  1.198(8)   &  1.469(15)  &  1.643(109)  &  1.476(34) \\
0.1266 &  1.153(9)   &  1.429(17)  &  1.604(107)  &  1.432(41) \\
0.1273 &  1.101(12)  &  1.385(21)  &  1.561(106)  &  1.387(54) \\
0.1279 &  1.057(15)  &  1.353(27)  &  1.530(109)  &  1.351(76) \\
0.1286 &  1.006(22)  &  1.331(43)  &  1.502(119)  &  1.301(126)
\end{tabular}
\end{ruledtabular}
\end{center}
\end{table}

\begin{figure}
\begin{center}
\includegraphics[height=0.85\hsize,angle=90]{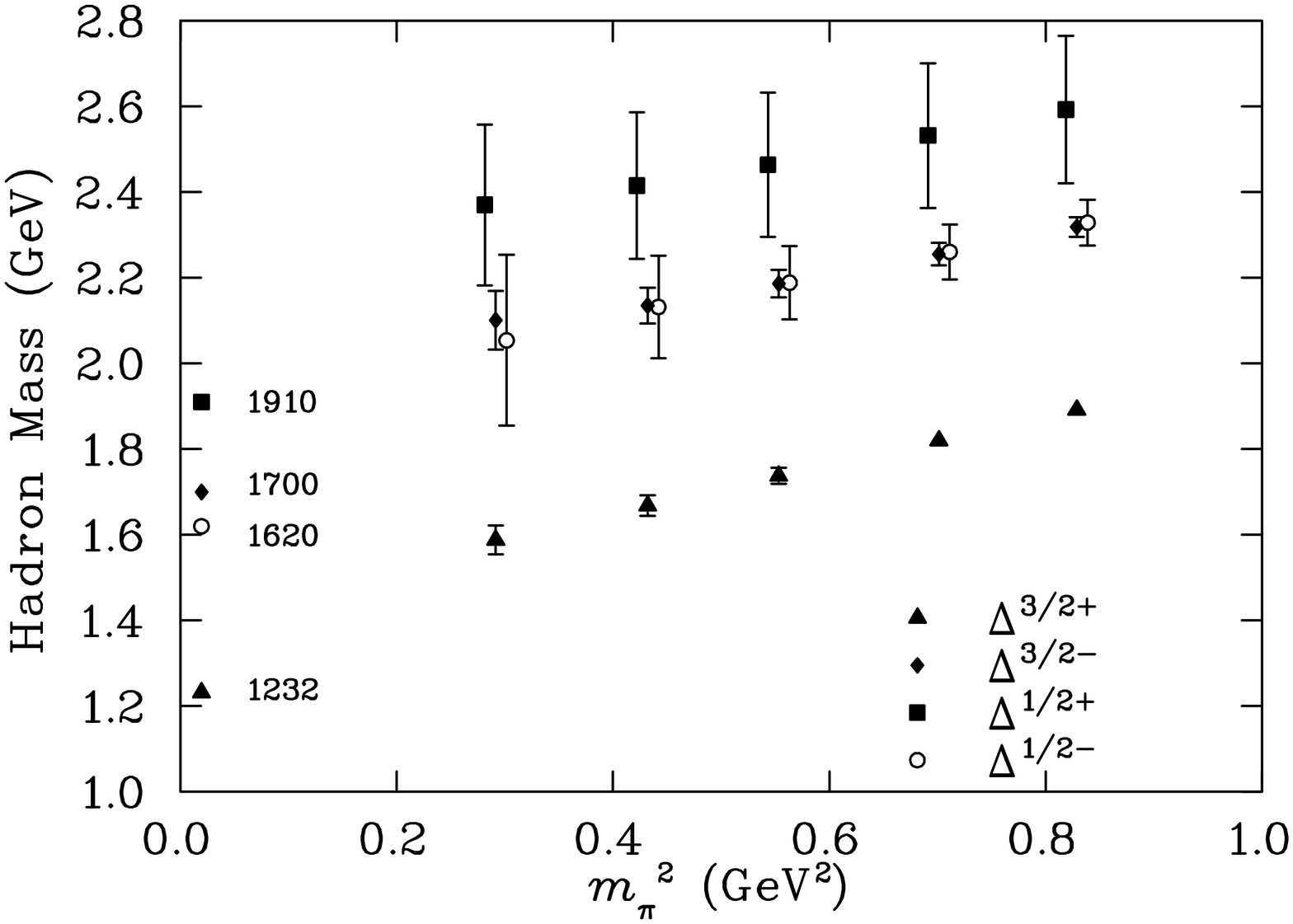} 
\vspace*{0.5cm}
\caption{Masses of the spin-projected $\Delta{3\over 2}^\pm$ and
  $\Delta{1\over 2}^\pm$ resonances.  The empirical values of the
  masses of the $\Delta^{3/2^+}(1232)$, $\Delta{3\over 2}^-(1700)$,
  $\Delta{1\over 2}^-(1620)$ and $\Delta{1\over 2}^+(1910)$ are
  shown on the left-hand-side at the physical pion mass.
  \label{Dpi2}}
\end{center}
\end{figure}

The results for the $\Delta{3\over 2}^+$ and $\Delta{3\over 2}^-$
masses are shown in Fig.~\ref{Dpi2} as a function of $m_\pi^2$.
The trend of the $\Delta{3\over 2}^+$ data points with decreasing $m_q$
is clearly towards the $\Delta(1232)$, although some nonlinearity with
$m_\pi^2$ is expected near the chiral limit \cite{MASSEXTR,ROSS}.
The mass of the $\Delta{3\over 2}^-$ lies some 500~MeV above that of
its parity partner, although with somewhat larger errors, as expected
from the effective mass plots in Figs.~\ref{D32+Mass} and \ref{D32-Mass}.

After performing a spin projection to extract the $\Delta{1\over
  2}^\pm$ states, a discernible, but noisy, signal is detected. This
indicates that the interpolating field in Eq.~(\ref{deltaIF}) has only
a small overlap with spin-$\frac{1}{2}$ states.
However, with 392 configurations we are able to extract a mass for the
spin-$\frac{1}{2}$ states at early times, shown in
Fig.~\ref{Dpi2}. Here we see the larger error bars associated with the
$\Delta{1\over 2}^\pm$ states.
The lowest excitation of the ground state, namely the $\Delta{1\over
  2}^-$, has a mass $\sim 350$--400~MeV above the $\Delta{3\over
  2}^+$, with the $\Delta{3\over 2}^-$ possibly appearing heavier.
The $\Delta{1\over 2}^+$ state is found to lie $\sim 100$--200~MeV
above these, although the signal becomes weak at smaller quark masses.
This level ordering is consistent with that observed in the empirical
mass spectrum, which is also shown in the figure.

The $N{1\over 2}^-$ and $\Delta{1\over 2}^-$ states will decay to
$N\pi$ in $S$-wave even 
in the quenched approximation \cite{QQCD}. For all quark masses
considered here, with the possible exception of the lightest quark,
this decay channel is closed for the nucleon.
While there may be some spectral strength in the decay mode,
we are unable to separate it from the resonant spectral strength.

The $N{3\over 2}^+$ and $\Delta{1\over 2}^+$ states will decay to
$N\pi$ in $P$ wave, while $N{3\over 2}^-$ and $\Delta{3\over 2}^-$
states will decay to $N\pi$ in $D$-wave.
Since the decay products of each of these states must then have equal
and opposite momentum and energy given by
$$
E^2 = M^2 + \left( \frac{2\pi}{aL} \right)^2 ,
$$
these states are stable in our calculations.

\section{Conclusion}
\label{Spin32Conclusion}

We have presented the first results for the spectrum of spin-$3\over 2$
baryons in the isospin-$1\over 2$ and $3\over 2$ channels, using a novel
Fat Link Irrelevant Clover (FLIC) quark action and an ${\cal O}(a^2)$
improved gauge action.
Clear signals are obtained for both the
spin-projected $N{3\over 2}^\pm$ and $N{1\over 2}^\pm$ states from a 
spin-$3\over 2$ interpolating field.
In particular, the ${1\over 2}^\pm$ states are in good agreement with earlier
simulations of the nucleon mass and its parity partner using the standard
spin-$1\over 2$ interpolating field. We find the $N{3\over 2}^-$
state to lie at a similar energy level to the $N{1\over 2}^-$,
consistent with experiment.
We also find a mass difference of $\sim 300$~MeV between the
spin-$\frac{3}{2}$, isospin-$\frac{1}{2}$ parity partners, slightly
larger than the experimentally observed difference of 200~MeV.

For isospin-${3\over 2}$ baryons, 
good agreement is found with earlier calculations for the $\Delta$ ground
state, and clear mass splittings between the ground state and its parity
partner are observed after suitable spin and parity projections.
We obtain a signal for the $\Delta{1\over 2}^\pm$ states and
the level ordering is consistent with that observed in the empirical
mass spectrum.

It will also be important in future work to consider the excited
states in each $J^P$ channel, in particular the lowest ``Roper-like''
excitation of the $\Delta(1232)$ ground state.
Although this will be more challenging, it may reveal further insights
about the origin of the inter-quark forces and the nature of the
confining potential.

\begin{acknowledgements}
  
  W.M. would like to thank R.G.~Edwards and D.G.~Richards for helpful
  discussions.  This work was supported by the Australian Research
  Council, and the U.S. Department of Energy contract
  \mbox{DE-AC05-84ER40150}, under which the Southeastern Universities
  Research Association (SURA) operates the Thomas Jefferson National
  Accelerator Facility (Jefferson Lab). The work of S.C. was supported
  by the Korean Ministry of Education under the BK21 program.

\end{acknowledgements}





\end{document}